# Past performance, peer review, and project selection: A case study in the social and behavioral sciences[1]

Peter van den Besselaar[2] & Loet Leydesdorff[3]

**Abstract**
Does past performance influence success in grant applications? In this study we test whether the grant allocation decisions of the Netherlands Research Council for the Economic and Social Sciences correlate with the past performances of the applicants in terms of publications and citations, and with the results of the peer review process organized by the Council. We show that the Council is successful in distinguishing grant applicants with above-average performance from those with below-average performance, but within the former group no correlation could be found between past performance and receiving a grant. When comparing the best performing researchers who were denied funding with the group of researchers who received it, the rejected researchers significantly outperformed the funded ones. Furthermore, the best rejected proposals score on average as high on the outcomes of the peer review process as the accepted proposals. Finally, we found that the Council under study successfully corrected for gender effects during the selection process. We explain why these findings may be more general than for this case only. However, if research councils are not able to select the 'best' researchers, perhaps they should reconsider their mission. In a final section with policy implications, we discuss the role of research councils at the level of the science system in terms of variation, innovation, and quality control.

## 1. Introduction

Research councils organize selection processes based on a variety of grant proposals. This variety is generated by creative researchers who translate new ideas into project proposals for which they try to secure funding. The aim of the councils is generally to select the "best" and most innovative proposals and researchers. In this report we address the question of how this process of variation (by the researchers) and selection (by the research councils) works: are research councils in a position to select the "best" proposals of the "best" researchers for funding? Is a council able to "pick the winners" on the basis of reviews? (Martin & Irvine, 1984).

1 We acknowledge the collaboration and support of the board and staff of the Netherlands Economics and Social Science Research Council MAGW-NWO (Van den Besselaar & Leydesdorff 2007). The paper also benefitted from discussions at the WZB workshop on peer review reviewed, Berlin, April 24-25, 2008,and the comments of three anonymous referees.

2 Peter van den Besselaar, Science System Assessment Department, Rathenau Instituut, Anna van Saxenlaan 51, 2593 HW Den Haag, Email: peter@rathenau.nl & Department of Organization Sciences, Vrije Universiteit Amsterdam.

3 Loet Leydesdorff, Amsterdam School of Communications Research (ASCoR), University of Amsterdam, Kloveniersburgwal 48, 1012 CX Amsterdam; email: loet@leydesdorff.net; http://www.leydesdorff.net .

This study is based on an analysis in collaboration with the Netherlands Economic and Social Science Research Council (MAGW-NWO). We were invited to evaluate their existing system of peer-review-based grant allocations. Would a scientometric evaluation of the proposals validate the results of their selection process? Can systematic differences between the two methods of evaluation be explained? How might this comparison change one's assessment of the value of both evaluation techniques in the social sciences?

Our primary research question focuses on the relationship between the past performance of researchers and the allocation decisions of the Research Council: does the money flow to the top *researchers* according to their past performance? Our second question addresses the relationship between the peer review system used by this Council and the decisions made by it: are the most highly rated *proposals* rewarded? As a third question, we focus on the relationship between the peer review reports and the past performance scores of the applicants.

Both peer review and the bibliometric assessment of performance have often been discussed and researched, and the two methods are often contrasted (Aksnes & Taxt 2004; Butler 2007, 2008; Charlton & Andras 2007; Donovan 2007; Moed 2007; Nightingale & Scott 2007). Peer review occurs in the context of journals selecting papers for publication as well as grant allocations. Reviewers are known to be inconsistent in their assessments of proposals and papers (e.g., Rothwell & Martyn 2000). The debate about peer review was triggered by an experiment in which already published papers were resubmitted, but with fake names and low-status universities as addresses. The resubmitted papers were mostly rejected, and only in a few cases did the reviewers recognize them as previously published (Peters & Ceci 1982). Some authors argue that results of this nature show that peer review is not an adequate selection process. Scientists, however, consider peer review as a necessity in an open research environment. From a discursive perspective, disagreement between assessments by referees can even be appreciated positively (Hirschhauer 2004).

Furthermore, specific problems may originate from the method of peer review itself. Some researchers observed a strong gender bias and nepotism in peer review (Wenneras & Wold, 1995), although others were not able to replicate this bias in more recent years (Sandstrom & Hällsten 2005). In a meta-analysis, Bornmann *et al.* (2008b) confirmed the continuing existence of a gender bias, but this bias seems to be diminishing. Nepotism, on the other hand, has remained an issue (Sandstrom &

Hällsten 2008). Another problem is the assessment of interdisciplinary research and research proposals. Laudel *et al.* (2006), for example, suggested that interdisciplinary research receives lower grades in peer review processes.

Peer review comes in a variety of formats, and often it is actually more like a committee review (Langfeldt 2001). In such a selection process, committee members provide peer assessments to a small subset of the submitted applications, but as committee members they advise or decide about all applications. Strategic behavior may then prevail; committee members often reach compromises so that each member warrants funding for one's favorite proposals (Langfeldt 2004).

The critique of peer review has resulted in a variety of proposals to change the procedures (Frolich 2003; British Academy 2007). However, a number of studies suggest that peer review works quite well. For example, Bornmann & Daniel (2006a, b) studied the selection of grant applicants extensively, using the data of a well-known and prestigious funding organization. The criteria used for these assessments were reliability, validity, and justice. In these studies, the conclusion was that despite a percentage of Type I and Type II errors, peer review works well on average.

Criticism of peer review has resulted in proposals to use bibliometric quality indicators as a more objective alternative (Irvine & Martin, 1983). In the UK, the recent changes in the Research Assessment Exercise (RAE) also point in this direction. In 2008, the RAE will be accompanied experimentally with a bibliometric evaluation in certain fields (Barker 2007; Oppenheim 2008). This has been criticized, mainly because bibliometric indicators are considered less applicable in some fields than in others, and because existing bibliographic databases are incomplete and biased, e.g., towards publications in English. Furthermore, citation patterns are different in different (sub)disciplines, and, therefore, bibliometrics indicators are difficult to compare across research fields (Leydesdorff, 2008).

At national and international levels, efforts are currently under way to shape the humanities and social sciences so as to enable bibliometric assessment in the future, for example, by defining lists of international top journals for the respective research specialties.[4] Another criticism, however, is that bibliometric statistics are hardly applicable at the level of individual researchers or applications, and above all, that the meaning of the indicators is still unclear. Do citations indicate quality or visibility? Furthermore, perverse effects are signaled: researchers would no longer focus on interesting research, but on adapting to the evaluation criteria (Wessely 1998; ESF 2006; Lawrence 2007; Adler et al 2008).

---

[4] Interestingly, the editors of many journals in science and technology studies and in the history of science have objected strongly this development in an editorial recently published in these journals (Andersen *et al.*, 2009).

While reviewing both the academic debate and the policy debate, we have found that there is not much agreement on how to organize the evaluation process. Results point in different directions with respect to both peer review and evaluations based on bibliometric assessments. In the policy debate, policy and managerial actors tend to prefer objective (bibliometric) indicators, whereas representatives of research communities emphasize that peer review may not be perfect, but remains the best available method (Wessely 1998). Opinions seem more to reflect positions than the processes under study.

**2. The case**

MaGW-NWO is the research council for the social and behavioral sciences in the Netherlands. It covers all social and behavioral disciplines: economics, management, psychology & pedagogy, political science & public administration, sociology, anthropology, communication studies, geography, demography, and law. The Council distributes research funds among researchers, institutes, and infrastructures, and increasingly plays a leading role in these domains in agenda setting, coordination, and network formation.

The Council has a budget of approximately 36 M€/year, of which 8 M€ is used for an open competition, 10 M€ for career grants, and 13 M€ for thematic research grants. In this study we assess the outcomes of the open competition and the career grants; these make up together about half of the Council's budget. The thematic grants were excluded since one can expect factors other than scientific excellence to play a role in this funding, such as the expected societal benefits of the research.

We were given access to applications for the years 2003, 2004, and 2005. Our dataset consists of 1178 applications, distributed unevenly over the disciplines: from 347 in psychology, 274 in economics, and 206 in law to 30 in communication studies, 28 in anthropology, and 9 in demography. Of these applications, 275 were accepted for funding. The applications cover four different funding programs: the open competition (629, of which 155 were granted), the young researchers' career program ("Veni": 428, of which 65 were granted), a second career development program ("Vidi": 100, of which 43 were granted) and finally the top career program ("Vici": 29 applications, of which 12 were granted).

The review process is organized differently in the various programs. In the open competition, proposals are reviewed by peers, and the researchers are given an opportunity to react to their comments. The proposal with a CV, the reviews, and the reply are input for the Council that makes the decision. In case of the career grants, the procedures vary. In the program for young researchers' grants, a pre-selection process takes place within a committee before the proposals are sent out for review. After the review, a selection of researchers is invited to present their proposals to the committee. The decision of the Council is based on the review scores, the proposal, the CV of the researcher, and this presentation.

## 3. The model

We use a model (Figure 1) to explain the success of a grant application in terms of the factors introduced above. In this schematic representation of the relevant processes, the decision about research funding is considered to be based on the quality of (1) the researcher involved, (2) the proposal, and (3) the network of the applicant. The quality of the proposal is operationalized as the judgments of peers, which are based on a review process. The past performance of the applicant (4) and his/her network (5) may play a role in this assessment.[5]

Various studies have indicated the existence of 'sexism and nepotism in science' (e.g., Wenneras & Wold, 1998). Furthermore, the interests of the actors involved and other social factors should be taken into account. Several variables can be considered in this context, such as the gender of the applicants, or network relations between applicants, reviewers, and decision makers which may partly be based on shared disciplinary and university affiliations. Each of these factors may influence the decisions of the research council (6), but also the comments of the referees (7).

The focus of this study is on the effects of past performance (arrow 1 in Figure 1), the referee's assessments (arrow 2), and the quality of the applicant's network (arrow 3) on the funding decision. Additionally, we test the extent to which the review outcome is related to past performance (arrow 4) and the applicants' network (arrow 5). Finally, we analyze how contextual factors play a role (arrows 6 and 7), as this may inform us about the quality or bias of the procedures of the Council. We did not measure the effect of grant allocations on post-performance, as this would be too early in reference to the data that were used in this study (that is, applications to the research Council in 2003, 2004 and 2005).

---

5 In case of the young researchers program, expected future performance may enter as a consideration more importantly than past performance.

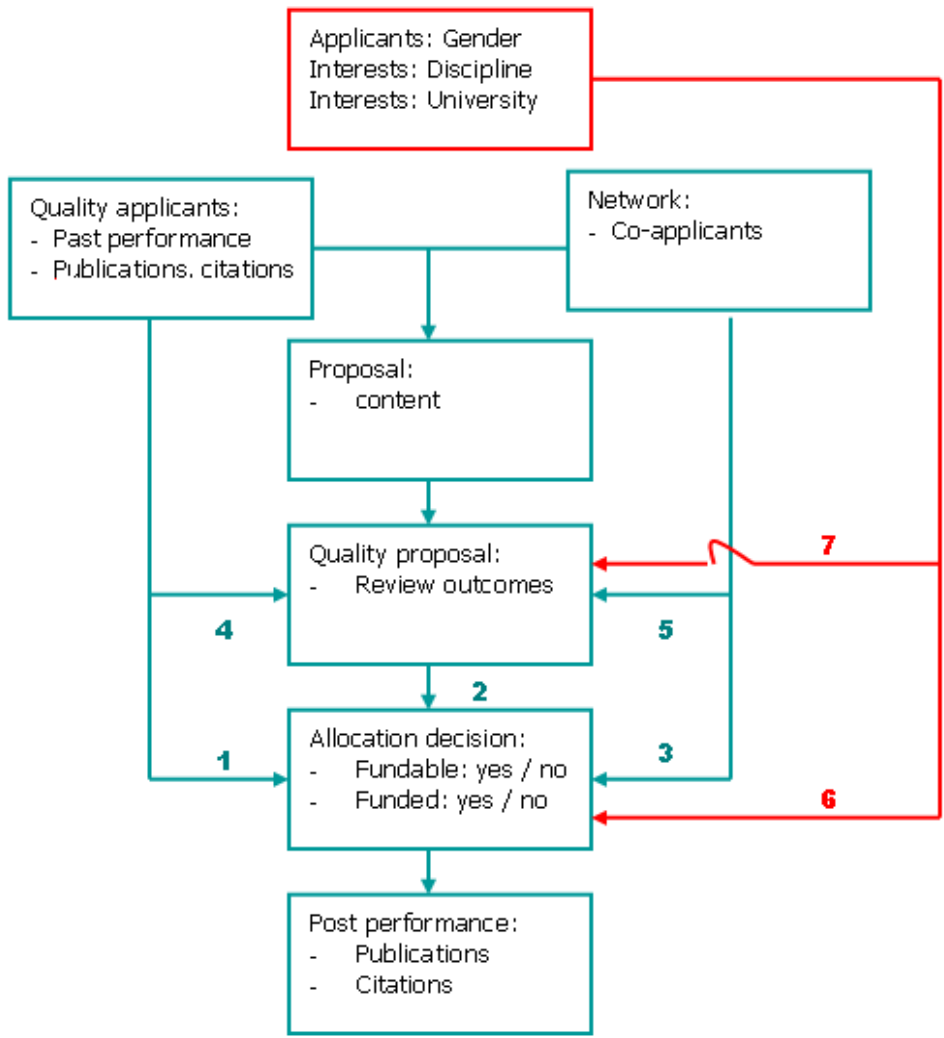

**FIGURE 1:** Schematic representation of the model.

## 4. Data and methods

The applications are our units of analysis. The *dependent variable* is the decision about the application, and this variable has three values: 'fundable' and funded (A); 'fundable' but not funded (A-); and 'non-fundable' applications (B). The applications vary in a number of respects.

- Firstly, we obtained data covering three years: 2003, 2004, and 2005. On some occasions, it will be relevant to compare the results across years.
- Secondly, the data cover four different *instruments.* The Research Council uses a variety of funding instruments, and the data accordingly distinguish between personal grants for researchers in three different phases of their careers and the open competition (OC).[6] The Council emphasized that the role of past performance may differ per instrument; hence there is also a need to distinguish among the instruments. In particular, the personal grants for young researchers (the so called VENI awards) may not be based on past performance, since these researchers are probably too young to have a strong track record. We therefore analyzed the data for the four instruments separately.
- Thirdly, the applications cover different disciplines, and past performance indicators (e.g., citation behavior and the orientation on *SoSCI* journals) can be expected to differ across disciplines and specialties. Our analysis starts at the aggregate level of the social and behavioral sciences as a whole, but we also test the model at the level of the disciplines and sub-disciplines.

Our first independent variable is *past performance.* Since this is a case study about a social science research council, data from the *Social Science Citation Index* (*SoSCI*) could be used as a representation of past performance of the relevant communities. For this measurement, we agreed with the Council to use the number of publications and citations received by both applicants and non-applicants (with a Dutch address) in the three year period prior to their application. We agreed with the Research Council to use this three-years period because the application form provides little space for the applicant to list his/her (recent) publications. In other words, success in 2003 is related to the publication data of 2001-2003 and the citations to these publications, success in 2004 to performance data of 2002-2004, and success in 2005 to performance data of 2003-2005.[7]

The number of publications in a certain year is defined in terms of publication years as indicated on the publications. (These dates are sometimes delayed when compared with calendar years.) The "times cited" are measured on February 9, 2007. All articles, reviews, notes, and letters with a Dutch address are included. While the applications to MaGW provide our units of analysis, a single researcher who applies more than once in a single year, may score differently on the grants received, but not on these performance indicators. If a researcher applies in different years, however, past performance can be expected to vary among years.

---

6 The thematic programs were not included in this study. Nevertheless, it would be useful to analyze these too, as the goals of the thematic programs are different and include criteria such as the possible societal outcomes of the research. These criteria are more opaque than the ones we analyzed in this study.

7 The performance of the co-applicants was measured in the same way.

This citation window is disputed within the social sciences (SWR-RGW 2005), but we were able to agree upon this procedure with the Research Council. A few issues are relevant to this decision:

- First, in some of the social sciences, books and not journals are the main publication outlets. We did not take these other publication formats into account, and for several fields (such as law) this is a serious disadvantage. For these fields, the current analysis should be interpreted with even more care (Nederhof 2006).[8]
- Second, the journals indexed in the *SoSCI* do not cover the complete relevant journal space. Furthermore, the coverage of the *SoSCI* varies for different parts of the social sciences. The orientation on publishing in journals which are processed by the Institute of Scientific Information (ISI) for inclusion in the *SoSCI* also differs per subfield (Van den Besselaar 2007).
- Third, even if sub-fields of the social sciences were covered equally well by the *SoSCI*, the disciplines and specialties are not homogenous in terms of publication and citation patterns.

Despite these shortcomings, the operationalization of quality in terms of citations and publications in ISI-indexed journals is increasingly institutionalized in Dutch universities; for example, in the case of the evaluation of graduate schools. Although many researchers have doubts about the validity of these indicators, the funding and evaluation procedures increasingly use them. Given these institutional incentives, researchers themselves increasingly try to publish in the 'top journals' as defined by impact factors.

The last name and the first initial were used to match the ISI data and the application data. This generates some error. For example, one of Van den Besselaar's articles is listed in the ISI-database as authored by "Van den BesselaarA", and L. Leydesdorff also appeared as "T Leydesdorff" in the database. These errors were not (manually) corrected. Our results should therefore be read as statistics with margins of error.

---

[8] As all these disciplines are evaluated in the same procedure by the Council, we start by including all disciplines. We will of course check if this influences the outcomes of our analysis, and repeat the analyses below also for the disciplines separately.

Our second independent variable is the *quality of the proposal* in the opinion of the referees. The data provided by the Research Council included the scores of the referees. However, the different instruments use a different classification system, and the classification schemes have changed slightly over the years. Two of the three scales were five-point scales, the other a three-point scale. We translated these different measures into a single five-point scale, using the scheme shown in Table 1.

**TABLE 1:** Coding of the referees' scores[9]

| Coding | OC | OC / VI | VI* |
|---|---|---|---|
| 1 | A | Excellent | Continue |
| 2 |   | Very Good | Doubt/continue |
| 3 | B | Good | Doubt |
| 4 |   | Fair | Doubt/stop |
| 5 | C | Poor | Stop |

OC: open competition;
VI: 'vernieuwingsimpuls' (three 'career grant' programs)
VI*: used in the pre-selection phase of the 'vernieuwingsimpuls'

The third independent variable is the *quality of the network* of the applicant. This variable is operationalized as the quality of the co-applicants in terms of past performance in the same way as the past performance of the applicants—that is, using numbers of publications and citations. We used two different operationalizations to calculate indicators for the quality of the network:

- Average network quality: average number of publications by the applicant and co-applicants, and average number of citations received by the applicant and co-applicants;
- Maximal network quality: the number of publications of the most productive applicant or co-applicant, and the number of citations received by the most highly cited applicant or co-applicant.

Only the applications for the open competition can be included in this last analysis, as career programs do not involve co-applicants. One may argue that other dimensions of an applicant's network could also be considered, such as the status of the applicant's

---

[9] It may seems strange to put "very good" and "doubt/continue" into one category? However, this reflects the Dutch research evaluation context. A score of 3 (good) is considered as low (doubtful quality), and one is generally unsatisfied with a score lower than 4.5 (in between very good and excellent)

PhD supervisor, the co-author network of the applicant, etc. In our opinion, these are interesting topics for further research.

Several *methods* were used for the analysis. We compare the group of successful with the group of unsuccessful applications using ANOVA: do these two groups differ significantly in terms of the independent variables? Using correlation analysis, we relate the past performance of the applicants and the review scores of the applications (the independent variables) with the amount of funding received. Using discriminant analysis, we test whether the independent variables could be used to 'predict' whether an application would be successful or not. In the first and the last analysis, the amount of the funding is not taken into account, but only the dichotomous variable of whether an application is funded or not.

**TABLE 2:** Variables of this study; the units of analysis are 1186 applications.

| | |
|---|---|
| Pub | Number of publications by the applicant (three years before the application) |
| Cit | Number of citations to these publications at 7 February 2007 |
| Pub2 | Average number of publications by applicant and the co-applicants |
| Cit2 | Average number of citations to the publications of the (co-)applicants, at 7 Feb. 2007 |
| Pub3 | Number of publications by the most productive of the applicant and the co-applicants |
| Cit3 | Number of citations to the publications of the most cited of the (co-)applicants, at 7 Feb. 2007 |
| Sex | Gender of main applicant |
| Uni | University of main applicant |
| Disc | Discipline of application |
| Instr | funding instruments (three types of personal grants and open competition) |
| Ref | Average of the referee's reports |
| Dec | 'Fundability': assessment by research Council |
| Euro | 'Funding': grants received from the research Council in euros . |

## 5. Research questions

We posed the following questions in this study:

1. Are the A (funded), A- (fundable, unfunded), and B (unfunded) applicants different in terms of their past performance and average referee scores?

2. Do past performance and the referee scores (of these three groups) correlate with the funding received?

3. Can one predict the success of applicants from their past performance and/or the referee scores?

After answering these questions, we will analyze the influence of some mediating variables, such as differences in subfields, funding instruments, the quality of the co-applicants, and the gender of the main applicants. More specifically, we discuss the following two issues:

4. Are there differences between disciplines: are the results different in e.g., law, economics, psychology, and the other disciplines under study here? Does the discipline influence the relations between the performance variables and the probability of success?

5. And what about gender differences?

## 6. Peer review, performance, and successful applications

ANOVA (analysis of variance) can be used to test whether grant recipients score more highly in terms of numbers of publications, citations, and referee scores than the researchers who did not get their proposals accepted for funding. Table 3 provides the results of the ANOVA. The successful applicants publish significantly more than the failed applicants (4.5 versus 2.7 publications), and are significantly more cited (36.0 versus 15.6 citations). Finally, the referees are significantly more positive about the funded applications than the non-funded ones. The difference is one point on a five-point scale. The successful applications score 1.6 ('very good/excellent') and the non-funded applications score on average 2.7 (slightly better than 'good').

**TABLE 3:** Publications and citations by success (2003-2005)

|   |   | N | Mean | Std. Deviation | Std. Error | 95% Confidence Interval for Mean | | Minimum | Maximum |
|---|---|---|---|---|---|---|---|---|---|
|   |   |   |   |   |   | Lower Bound | Upper Bound |   |   |
| Citations | A | 275 | 36.03 | 70.704 | 4.264 | 27.64 | 44.43 | 0 | 593 |
|   | A- / B | 911 | 15.61 | 45.295 | 1.501 | 12.67 | 18.56 | 0 | 621 |
|   | Total | 1186 | 20.35 | 52.969 | 1.538 | 17.33 | 23.36 | 0 | 621 |
| Publications | A | 275 | 4.45 | 5.988 | .361 | 3.74 | 5.16 | 0 | 43 |
|   | A- / B | 911 | 2.71 | 4.915 | .163 | 2.39 | 3.03 | 0 | 62 |
|   | Total | 1186 | 3.11 | 5.233 | .152 | 2.81 | 3.41 | 0 | 62 |
| Referee | A | 274 | 1.59 | 0.634 | .036 | 1.52 | 1.66 | 1.00 | 3.67 |
|   | A-/ B | 904 | 2.68 | 1.045 | .035 | 2.61 | 2.75 | 1.00 | 5.00 |
|   | Total | 1183 | 2.43 | 1.064 | .031 | 2.36 | 2.49 | 1.00 | 5.00 |

|   |   | Sum of Squares | df | Mean Square | F | Sig. |
|---|---|---|---|---|---|---|
| Citations | Between Groups | 88091.424 | 1 | 88091.424 | 32.224 | .000 |
|   | Within Groups | 3236713.147 | 1184 | 2733.710 |   |   |
|   | Total | 3324804.571 | 1185 |   |   |   |
| Publications | Between Groups | 638.992 | 1 | 638.992 | 23.785 | .000 |
|   | Within Groups | 31808.317 | 1184 | 26.865 |   |   |
|   | Total | 1332.180 | 1177 |   |   |   |
| Referee | Between Groups | 247.143 | 1 | 247.143 | 267.862 | .000 |
|   | Within Groups | 1085.037 | 1176 | .923 |   |   |
|   | Total | 1332.180 | 1177 |   |   |   |

A: funded; A-: fundable, not funded; B: not-fundable

We also grouped the applications differently: 'fundable applications' (A and A-) versus 'non-fundable applications' (B). Again, the 'fundable' applications score significantly better on all variables than the 'non-fundable' ones. The differences are smaller (but still statistically significant) than when comparing the funded (A) and the non-funded (A- and B) applications. When comparing A with A-, again the A applications score significantly higher than the A-.

### *The problem of skewed distributions*

Most statistical techniques used assume normal distributions in the data, although, for example, discriminant analysis is robustly against violation of this assumption. To help interpret the statistical results correctly, we present graphs of the distribution of publications as an example (Figure 2). The figure shows on the left side the funded applications and on the right side the rejected ones. As this figure shows, the data are skewly distributed. The rejected ones include a long tail on the right side.

Figure 2 suggests that in the top segments of the distributions the number of publications per applicant is not very different between the two (successful and unsuccessful) groups. Furthermore, one can see that the (short) tail of the successful group contains applicants with low scores. Similar patterns can be shown for the distribution of the citations and referee scores. In any case, if there are differences between the two groups (of awarded and rejected applicants), these differences are not so large that one can identify them from visual inspection of the graphs.

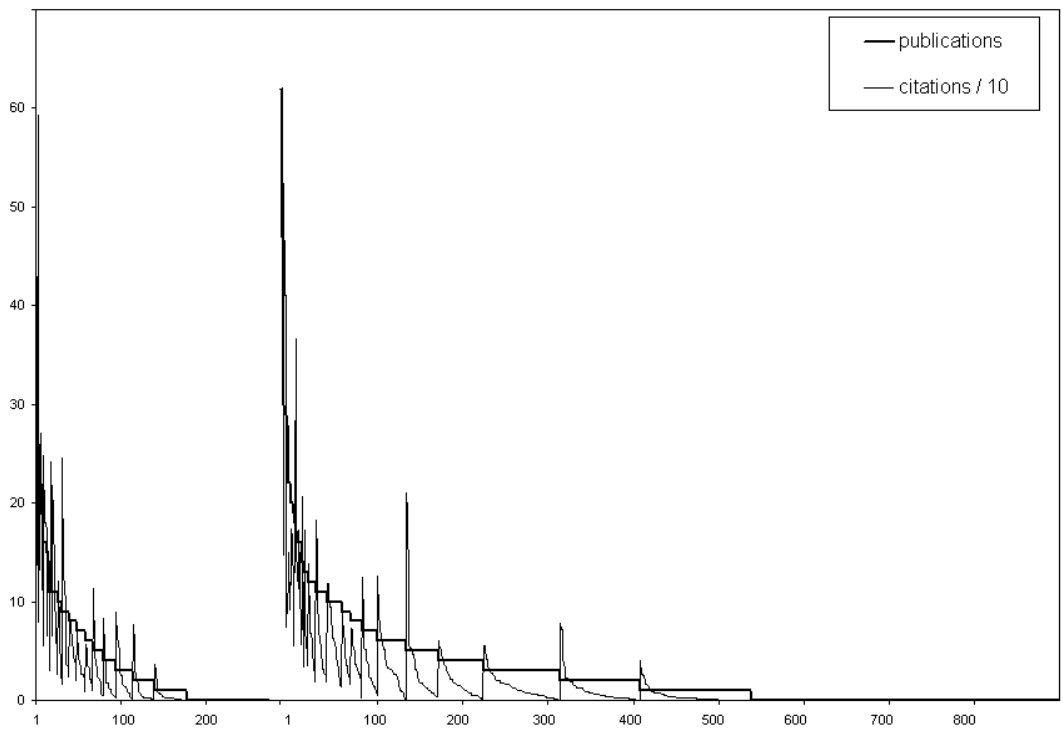

**FIGURE 2:** Distribution of publications; successful applicants (left) and unsuccessful (right)
(left axis: number of publications; number of citations/10)

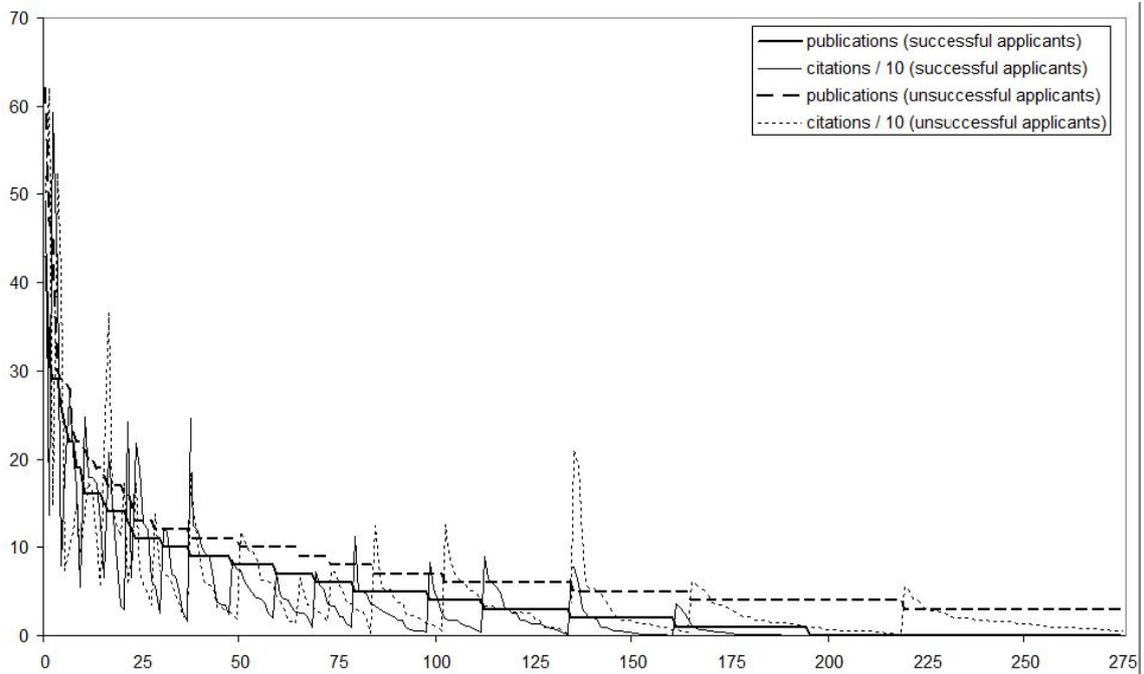

**FIGURE 3:** Superposition of the distribution of publications of successful and best unsuccessful applicants (left axis: number of publications; number of citations/100)

In order to obtain a more detailed overview, we plotted the distributions of publications of the 275 successful applicants and the *top* performing 275 non-successful applicants in a single graph (Figure 3).[10] The results are interesting because they show that the best scoring unsuccessful applicants score higher in terms of past performance than the successful applicants. In other words, the long tail of the lower performing applicants causes the differences in the averages between the two groups. We can conclude therefore that the skewed distributions in the data may indeed heavily influence the statistical results above.

When we restrict the ANOVA to the top 275 unsuccessful applicants and 275 successful applicants the results change radically.[11] As already suggested by Figure 3, unsuccessful applicants achieve a significantly higher score on the past performance indicator (Table 4). For the refereeing results, the initial difference disappears: in this smaller set no significant difference remains between the two groups.

**TABLE 4:** Publications and citations by success—top 550 only* (2003-2005)

|     |       | N   | Mean  | Std. Deviation | Std. Error | 95% Confidence Interval for Mean Lower Bound | 95% Confidence Interval for Mean Upper Bound | Min | Max |
|-----|-------|-----|-------|----------------|------------|----------------------------------------------|----------------------------------------------|-----|-----|
| pub | A     | 275 | 4.44  | 5.992          | .361       | 3.73                                         | 5.15                                         | 0   | 43  |
|     | A-/B* | 277 | 6.92  | 7.068          | .425       | 6.08                                         | 7.75                                         | 0   | 62  |
|     | Total | 552 | 5.68  | 6.664          | .284       | 5.13                                         | 6.24                                         | 0   | 62  |
| cit | A     | 275 | 36.03 | 70.708         | 4.264      | 27.63                                        | 44.42                                        | 0   | 593 |
|     | A-/B* | 277 | 48.04 | 72.338         | 4.346      | 39.48                                        | 56.59                                        | 0   | 621 |
|     | Total | 552 | 42.05 | 71.718         | 3.053      | 36.06                                        | 48.05                                        | 0   | 621 |
| ref | A     | 276 | 1.62  | .749           | .045       | 1.53                                         | 1.71                                         | 1   | 9   |
|     | A-/B* | 271 | 1.64  | .492           | .0300      | 1.58                                         | 1.70                                         | 1   | 3   |
|     | Total | 547 | 1.63  | .634           | .0277      | 1.58                                         | 1.68                                         | 1   | 9   |

|     |                | Sum of Squares | Df  | Mean Square | F      | Sig. |
|-----|----------------|----------------|-----|-------------|--------|------|
| pub | Between Groups | 846.670        | 1   | 846.670     | 19.711 | .000 |
|     | Within Groups  | 23624.850      | 550 | 42.954      |        |      |
|     | Total          | 24471.520      | 551 |             |        |      |
| cit | Between Groups | 19907.016      | 1   | 19907.016   | 3.891  | .049 |
|     | Within Groups  | 2814158.461    | 550 | 5116.652    |        |      |
|     | Total          | 2834065.476    | 551 |             |        |      |
| ref | Between Groups | .038           | 1   | .038        | .094   | .759 |
|     | Within Groups  | 219.418        | 545 | .403        |        |      |
|     | Total          | 219.455        | 546 |             |        |      |

A: funded; A-:fundable, not funded; B: not-fundable
*: stratified (by funding instrument) sample of best scoring unsuccessful applicants. "Best" is defined in two ways (past performance and referee score). This results in different groups.

With these results, *question one* of this study can be answered: *on average*, the funded applicants have a better past performance than the non-funded applicants, the fundable applicants have a better past performance than the non-fundable, and the funded applicants have a better past performance than the fundable-non-funded. However, this comparison is based on averaging over highly skewed distributions. The 275 non-funded applicants with the highest scores had significantly higher scores on past performance indicators than the 275 funded applicants. This suggests firstly that the selection was *not* based on past performance (as it was defined in this study), and secondly that the reservoir of potential recipients is much larger than the group that was awarded (Melin and Danell 2006).

---

10 Top performing is based on citations received and (secondly) on publications. Of these about 50% are considered as fundable by the Council, and the other 50% as non-fundable.

11 The distribution over the four instruments (OC: 154 and 159; Veni: 65 and 83; Vidi: 43 and 21, Vici: 12 and 12) is not equal in the two groups of successful and unsuccessful applicants. Correction for this – by using a stratified sample – does not change the results.

Similarly, the funded applicants have *on average* a better score in the referee reports than the non-funded applicants, the fundable applicants have on average a better referee's score than the non-fundable, and the funded applicants have on average a better referee's score than the fundable-non-funded (A-). But the 275 unsuccessful applicants with the highest referee scores perform at the same level as the successful applicants.

In summary, our results suggest that the system is successful in identifying and discarding the tail of the distribution according to the Mertonian norms of rewarding merit, but not successful in selecting among the applications by applicants who meet these norms.

## 7. Is one able to predict success from past performance?

Let us further analyze the relations between past performance, referee scores, and success in getting an application funded. The application is again the unit of analysis. Consequently, if researchers file more than one project, these projects are treated as two different cases. Because the data used are relatively skewed (publications, citations), an ordinal measure of association such as Spearman's rho is used.[12]

As Table 5 shows, past performance measures—numbers of citations received and numbers of publications—correlate strongly ($\rho = 0.92$). These performance indicators correlate positively with the amount of funding received, but this correlation is low ($\rho < 0.20$). There is also a low but significant correlation between referee's judgments and the past performance indicators ($\rho \approx 0.20$). Referee reports also correlate moderately with the funding received, but higher do than the past performance indicators ($\rho =$

---

12 The referees' judgments are less skewed, and the mean (2.42) and median (2.17) differ only slightly.

0.46). (The negative sign of these correlations in Table 5 can be read as positive; the inversion of the sign is caused by the scale used for measuring the referee's judgments, which rank applicants from one for 'excellent' to five for 'poor'.)

**TABLE 5:** Success by past performance and peer review

|  |  | Cit | referee | euro |
|---|---|---|---|---|
| Publications | Spearman's rho | .923 | -.205 | .160 |
|  | Sig. (2-tailed) | .000 | .000 | .000 |
|  | N | 1186 | 1178 | 1186 |
| Citations | Spearman's rho |  | -.214 | .185 |
|  | Sig. (2-tailed) |  | .000 | .000 |
|  | N |  | 1178 | 1186 |
| Referee | Spearman's rho |  |  | -.455 |
|  | Sig. (2-tailed) |  |  | .000 |
|  | N |  |  | 1178 |

*Discriminant Analysis* (DA) can be used to distinguish group membership (a nominal variable), such as successful and unsuccessful applications, from interval variables, such as the numbers of publications and citations, and the referee scores. We conducted several DAs, and all the analyses resulted in a significant model.

In the first DA we used 'publications' and 'citations' as independent variables. The model correctly classifies about 35% of the successful applications, and about 85% of the unsuccessful. This does not change in the stepwise model, which by the way removed 'publications' as a relevant predictor from the analysis. It shows that although successful and unsuccessful applications differ in terms of the numbers of publications by the applicants, these differences do not differentiate between the two classes. In other words, at this level of all instruments and all sub-disciplines, past performance (as defined in this specific way) is only weakly related to success.

We also used DA with 'publications,' 'citations,' and 'referee scores.' This improved the classifications considerably. In this case, the percentage of correctly classified successful applications is 85%, and the percentage of correctly classified unsuccessful applications is 61% (Table 6). A stepwise procedure does not change the result — but the 'publications' variable is again removed.

In summary, the reports of the referees contribute considerably to the correctly predicted classification. This accords with the results of the correlation analysis which showed above that the relation between 'funding received' and the 'referee's scores' is moderately high at 0.46. Nevertheless, a large part of the variance remains unexplained. In other words, the Council takes considerable responsibility for making its own funding decisions: it accepts applications with low referee scores, and it regularly rejects applications that received high appreciation from the reviewers.

**TABLE 6:** Classification of applications from pub, cit, and ref *

|  |  | A versus A- / B | Predicted Group Membership | | Total |
|---|---|---|---|---|---|
|  |  |  | A | A- or B |  |
| Original | Count | A | 231 | 43 | 274 |
|  |  | A- or B | 354 | 550 | 904 |
|  | % | A | 84.3 | 15.7 | 100.0 |
|  |  | A- or B | 39.2 | 60.8 | 100.0 |

a  66.3% of original grouped cases correctly classified.
* Stepwise: *cit* and *ref* remain in the analysis

### *The top 550 (2 x 275) again*

As in the previous analysis, the long tails of the unsuccessful group may heavily influence the outcome of the analysis. Therefore, we repeated the analysis including only the top segment of the unsuccessful applicants (stratified sample in terms of funding instruments). The results are striking: the correlations between past performance indicators and funding become negative, and these correlations are stronger than in the case of the whole sample (Table 7).

**TABLE 7.** Success by past performance*

|  |  | Cit | Euro |
|---|---|---|---|
| Pub | Spearman's rho | .833 | -.256 |
|  | Sig. (2-tailed) | .000 | .000 |
|  | N | 552 | 552 |
| Cit | Spearman's rho |  | -.262 |
|  | Sig. (2-tailed) |  | .000 |
|  | N |  | 552 |

*   Stratified (by funding instrument) sample of best scoring unsuccessful applicants

We applied the same stratification to the referees' scores. As Table 8 shows, the correlation between referee scores and awarded grants has disappeared—although it was rather large in the whole dataset.[13] In this case, none of the three independent variables explains success in obtaining funding.

---

[13] Since the definition of the unsuccessful applications with the highest scores is different from the one used in Table 7, correlations between 'publications,' 'citations,' and 'euro' differ between Tables 7 and 8.

**TABLE 8.** Success by review score and past performance*

|     |                   | Cit  | Euro | Referee |
|-----|-------------------|------|------|---------|
| Pub | Spearman's rho    | .917 | .086 | -.048   |
|     | Sig. (2-tailed)   | .000 | .155 | .266    |
|     | N                 | 547  | 276  | 547     |
| Cit | Spearman's rho    |      | .055 | .001    |
|     | Sig. (2-tailed)   | .    | .359 | .988    |
|     | N                 |      | 276  | 547     |
| Euro| Spearman's rho    |      |      | -.004   |
|     | Sig. (2-tailed)   |      |      | .950    |
|     | N                 |      |      | 276     |

\*   Stratified (by funding instrument) sample of best scoring unsuccessful applicants

The discriminant analysis works differently in this case than in the case of the whole sample: a larger number of the successful applications are now correctly classified. However, the number of correctly classified unsuccessful applications has decreased (Table 9). More importantly, the percentage of correctly classified cases is not much higher than 50%. This means that the prediction using the performance indicators as independent variables in the DA barely improves on a random classification.

**TABLE 9.** Classification of applications from pub en cit (stepwise*)

|          |       | decision | Predicted Group Membership | | Total |
|----------|-------|----------|------|------|-------|
|          |       |          | 1    | 2    |       |
| Original | Count | A        | 192  | 83   | 275   |
|          |       | A- or B  | 148  | 129  | 277   |
|          | %     | A        | 69.8 | 30.2 | 100.0 |
|          |       | A- of B  | 53.4 | 46.6 | 100.0 |

a  58.2% of original grouped cases correctly classified.
A = funded; A- = fundable, not funded; B = not funded

DA applied to the set including the unsuccessful applications with the best referee scores did not work. In this model, all variables were excluded from the discriminant analysis, indicating that publications, citations, and referee scores cannot be used to predict success within this sub-population.

We are now able to answer the *second* and the *third research questions* of this study: do past performance and success correlate? And, can one predict the success of applicants from their publication and citation scores?

- The correlation analysis shows a low correlation between the scientometric indicators of past performance and the amount of funding received. The correlations between past performance and referee scores are also low.

- If one wants to predict success and failure (not taking into account the amount of money received), the number of citations can be used statistically to predict group membership. However, only a low percentage of the successful cases are correctly classified.

- By including the referee scores, the percentage of correct predictions increases considerably.

- When comparing the group of successful and top 275 unsuccessful applicants, all relationships disappear, and in this case, we find a negative correlation between past performance and the amount of funding received. Between referee judgment and funding, the correlation disappears.

- We repeated the analysis at the *discipline level*, with the same results as for the whole set. Using ANOVA, in each of the disciplines the referee scores, the number of publications, and the citations received do not differ significantly between the successful and the best unsuccessful applicants.

Some additional tests were done:

- We repeated the analysis for the four different *funding instruments*, as the criteria and the selection procedures are different. The relations between the variables became even lower than for the set as a whole, with the exception of the VICI program. However, the *N* is very small in this case.

- We also studied the effect of the past performance of the co-applicants. The successful applicants have on average co-applicants with a higher past performance than the unsuccessful ones. Correlation between the network indicators and the outcomes of the review process are low, but higher than the correlation between the past performance of the main applicant and the results of the review process. The same holds for the correlation between these network indicators and the amount of funding.[14] We therefore may conclude that the quality of the co-applicants does positively influence the success of an application, but this effect is not strong.

---

[14] Including the network indicators in the discriminant analysis hardly improves the classification results.

## 8. Disciplinary differences?

As already stated, the variable 'past performance' may have a rather different meaning in the various fields. The relevant question then is whether or not the results obtained for the social sciences as a whole can be reproduced at the level of the disaggregated disciplines. Publication and citation behaviors vary among disciplines and fields.

Table 10 illustrates this difference for sociology and psychology journals as an example. Psychology journals (as defined by the ISI subject categories) have on average higher impact factors, and this indicates that psychologists have longer reference lists per paper than sociologists. Another reason could be that compared with psychologists, sociologists cite other document types (books, book chapters etc.) more frequently than journal articles. Furthermore, what counts as a top journal in psychology is different from what would count as a top journal in sociology, if one takes the impact factor of the journal (IF) as a criterion. The top 10% starts in sociology with an IF of 1.382, and in psychology with an IF of 3.458. The maximum value of the IF of sociology is even below this cutoff value with only 3.262.

**TABLE 10**: Differences in citation behavior between fields

|           | mean  | CoV*  | Median | max    | top 10%   | skewness | N    |
|-----------|-------|-------|--------|--------|-----------|----------|------|
| All SoSCI | 0.982 | 1.041 | 0.697  | 12.642 | IF >2.013 | 3.869    | 1745 |
| Psychology| 1.287 | 1.386 | 0.650  | 9.780  | IF >3.458 | 3.114    | 101  |
| Sociology | 0.683 | 0.879 | 0.460  | 3.262  | IF >1.382 | 1.982    | 94   |

*: coefficient of variance = standard deviation divided by mean; IF = Impact Factor

These differences at the disciplinary level are reflected in the scores on the variables used in this study. Indeed, the averages of the variables show a large variation among the disciplines (Table 11).

**TABLE 11:** Averages in the sample by discipline

|                  | N    | Pub[#] | cit  | Pub2 | Cit2 | Pub3 | Cit3 | ref | K€  |
|------------------|------|--------|------|------|------|------|------|-----|-----|
| Anthropology     | 28   | 1.4    | 5.3  | 1.0  | 3.8  | 1.4  | 5.3  | 2.4 | 234 |
| Communication    | 30   | 3.7    | 16.1 | 2.8  | 11.4 | 4.1  | 18.7 | 2.6 | 149 |
| Demography       | 9    | 2.3    | 7.0  | 3.1  | 11.8 | 4.6  | 20.2 | 2.4 | 288 |
| Economics**      | 274  | 2.4    | 8.5  | 2.4  | 8.8  | 3.0  | 11.4 | 2.6 | 235 |
| Education        | 79   | 2.8    | 19.7 | 2.9  | 20.1 | 3.3  | 21.9 | 2.8 | 180 |
| Geography        | 40   | 2.5    | 12.0 | 2.6  | 12.4 | 2.9  | 13.4 | 2.2 | 340 |
| Law              | 206  | 0.5    | 2.0  | 0.6  | 2.4  | 1.0  | 3.8  | 2.3 | 187 |
| Political sci*** | 56   | 1.2    | 3.8  | 1.2  | 4.0  | 1.6  | 5.9  | 2.5 | 209 |
| Psychology*      | 347  | 5.8    | 48.5 | 5.9  | 49.7 | 7.8  | 65.3 | 2.2 | 249 |
| Sociology        | 115  | 2.9    | 13.3 | 3.4  | 17.3 | 4.9  | 26.2 | 2.5 | 169 |
| Total            | 1184 | 3.1    | 20.3 | 3.2  | 21.1 | 4.2  | 27.9 | 2.4 | 226 |

[#] See table 1 for the variable names
* Incl. pedagogy; ** Incl. management; *** Incl. public policy

To find out if these differences between disciplines influence the findings of the previous sections, we repeated the analysis for the different disciplines separately. Table 12 shows the correlations between past performance, referee results, and the amount of funding for the disciplines individually. We omit the disciplines with N<50. On average, the indicators work better at the discipline level, as the correlations between past performance and referee scores are somewhat higher than at the aggregated level. The same holds for the correlation between past performance and funding received. However, the correlations are still not high. The relation between the review outcome and the received funding remains unchanged. However, a few cases deviate from these patterns.

- In law, the bibliometric indicators did not correlate at all with the review results and with the received funding. This is in line with the general opinion that research in the field of law is not oriented towards publication in international journals, but has other types of (mainly nationally oriented) output;
- In the case of political science, the bibliometric indicators did not correlate with the amount of funding;
- In the case of communication studies, the bibliometric indicators did not correlate with the referees' evaluations;
- In the case of demography, negative correlations were found, but these were not significant due to the low numbers ($N = 9$).

**TABLE 12:** Relations (Spearman's rho) between past performance, referee score and success by discipline

|  | PP~Rev | PP~Euro | Rev~Euro | N |
|---|---|---|---|---|
| All | 0.21 | 0.17 | 0.46 | 1181 |
| Average | 0.28 | 0.28 | 0.44 |  |
| Psychology* | 0.27 | 0.19 | 0.49 | 345 |
| Economics** | 0.33 | 0.15 | 0.42 | 274 |
| Law | - | - | 0.43 | 206 |
| Sociology | 0.17 | - | 0.45 | 114 |
| Education | 0.32 | 0.10 | 0.43 | 79 |
| Political science*** | 0.27 | - | - | 56 |
| Geography | 0.36 | 0.44 | 0.37 | 40 |
| Communication | - | 0.28 | 0.50 | 30 |
| Anthropology | 0.27 | 0.50 | 0.45 | 28 |
| Demography | - | - | - | 9 |

PP = past performance (average of variables pub and cit)
Netw = quality of the network (average of variables pub3 and cit3)
Rev = average of the referee scores (variable ref)
Euro = amount of money received (variable euro)
* Incl. pedagogy; ** Incl. management; *** incl. Incl. public policy

When we disaggregated to an even lower level, the analysis did not further improve. For example, we calculated the correlations between past performance, referee scores, and the success of the applications for the four sub-disciplines of psychology, as distinguished by the Research Council. Some of the correlation are higher than for psychology on average; others are very low and/or not significant (Table 13).

**TABLE 13:** Relations between past performance, referee score and success by sub-discipline of psychology

| (Spearman's rho) | PP~Rev | PP~€ | Rev~€ | N |
|---|---|---|---|---|
| All | 0.21 | 0.17 | 0.46 | 1181 |
| Psychology (incl. pedagogy) | 0.27 | 0.19 | 0.49 | 345 |
| Clinical /biological /medical psychology | 0.32 | | 0.45 | 85 |
| Developmental psychology & pedagogy | 0.25 | 0.35 | 0.49 | 68 |
| Cognitive and biological psychology | 0.28 | | 0.49 | 109 |
| Social, work and organizational psychology; psychometrics | 0.24 | 0.29 | 0.58 | 83 |

Finally, we used discriminant analysis at the level of disciplines. If we use only the past performance indicators in the analysis, the results at the disciplinary level are similar to those for the whole set, with the exception of law (83% correct positives, 15% correct negatives). But we have already seen that the indicators do not seem to work in the case of law (Table 12). Interestingly, if we use all the variables in a stepwise analysis, only the peer review variable is retained, and the past performance indicators are removed from the analysis.

We now can answer *question 4*: the indicators work slightly better at the level of disciplines, but this does not change the results obtained in the previous sections for the total set. In other words, the decisions at the disciplinary level are somewhat stronger related to past performance. The effects of the referee reports, however, did not change.

## 9. Contextual factors: discipline and gender

Does the discipline matter if one wants to obtain funding for a proposal? Table 14 shows two interesting patterns. First, the accepted applications are unevenly distributed over the disciplines. Three disciplines have a large share of all accepted proposals: psychology and pedagogy (43%), economics (17%), and law (14%). The remainder is only 26%. Furthermore, the acceptance rates differ considerably between the sub-disciplines, from 10% in political science to 32% in psychology.

**TABLE 14:** Accepted applications by subfield (OC and VI)

| | Number of applications | Applications By field | Accepted | Rejection Rate | Accepted by field | Professors* |
|---|---|---|---|---|---|---|
| Anthropology | 30 | 2.4 | 9 | 70.0 | 3.3 | 2.3 |
| Communication | 31 | 2.4 | 5 | 83.9 | 1.8 | 1.5 |
| Demography | 10 | 0.8 | 2 | 80.0 | 0.7 | 0.1 |
| Economics** | 301 | 23.7 | 46 | 84.7 | 16.8 | 35.6 |
| Education | 81 | 6.4 | 14 | 82.7 | 5.1 | 3.3 |
| Geography | 47 | 3.7 | 8 | 83.0 | 2.9 | 3.4 |
| Law | 219 | 17.2 | 41 | 81.3 | 15.0 | 28.4 |
| Political science*** | 61 | 4.8 | 6 | 90.2 | 2.2 | 6.0 |
| Psychology**** | 370 | 29.1 | 119 | 67.8 | 43.4 | 15.0 |
| Sociology | 121 | 9.5 | 24 | 80.2 | 8.8 | 4.5 |
| Total | 1271 | 100% | 274 | 78.4 | 100% | 100% |

\* Full and associate professors by field — general universities. Source: NOD, 2004.
\*\* Incl. management    \*\*\* Incl. public policy    \*\*\*\* Incl. pedagogy

Secondly, the distribution of successful applications over the disciplines reflects the size of the disciplines. The last column in Table 14 shows the relative sizes of the disciplines in the Netherlands universities, in terms of numbers of full professors and associate professors. The differences in size are rather large. The number of professors per discipline correlates highly ($r = 0.80$) with the number of applications, and moderately ($r = 0.53$) with the number of successful applications. In sum, 'redistribution' takes place among the large disciplines. Economics and law get substantially less than could be expected given their size. Psychology gets more: 15% of the senior staff produces 29% of the applications, and 46% of the successful ones.[15]

Does gender matter in getting a proposal accepted? In the dataset under study, about 32% of all applications have a female principal investigator. The activity of men and women in the various fields and instruments are different, as are their success rates. The question of gender bias in (peer) review procedures is an important issue (Wenneras & Wold 1997). Empirical research shows contradictory findings, although a recent meta-analysis suggests that a (relatively small) gender bias exists (Bornmann *et al.*, 2008b).

Let us define gender inequality in our model as men having a better chance than women to get a project funded, given the same referee results and with the same past performance. For example, if male researchers have a better past performance, this may explain the higher rating by the referees and it may explain the higher success rate of male researchers. Table 15 shows that male and female researchers indeed do differ significantly in terms of publications, citations, and referee results. Male researchers receive higher scores from referees. The average score of the female researchers is 87% of the score of male researchers. Female researchers also score lower on past performance indicators (about two thirds) than do male researchers. If we distinguish among the three groups of A, A- and B applications, the picture becomes a bit different: in all cases male applicants score better on the three predicators, but for the A and A- applications, the difference between the number of citations of male and female researchers is no longer significant.

---

15 The dominance of psychology is even stronger on the level of the individual schemes: 75% of the Vici grants – the most prestigious one among the four included in this study – are for researchers in the field of psychology.

Does the selection process of the Research Council shows a gender bias? As Table 15 shows, female researchers do not seem to be disadvantaged by the *refereeing process* since the gender differences are smaller in this dimension than in terms of past performance measurements. Furthermore, although successful male and female applicants receive about the same number of citations, the differences between the number of publications and the scores by the referees remain significant (at 10%).

**TABLE 15:** Performance by gender

|  | Average nr of publications | Average nr of citations | Average referee score | N |
|---|---|---|---|---|
| Male (all) | 3.52 | 23 | 2.32 | 808 (68.1%) |
| Female (all) | 2.24 | 16 | 2.66 | 378 (31.9%) |
| Female/male | .64 | .69 | .87 |  |
| Male (granted) | 4.70 | 37 | 1.55 | 194 (70.5%) |
| Female A (granted) | 3.84 | 34 | 1.71 | 81 (29.5%) |
| Female/male | .82 | .92 | .91 |  |

Finally, we compared the shares of female researchers in the 275 best reviewed applications, in the set of the 275 most publishing applicants, and in the set of the 275 most cited applicants (Table 16). In all three 'top lists' the shares of female researchers are smaller than the share of women in the set of successful applicants. This result suggests that during the final decision-making process the Council corrects the results of the review process in favor of female applicants — reflecting a deliberate policy of the Council to stimulate women to pursue research careers.

**TABLE 16:** Gender bias?

|  | Male | Female | % Female |
|---|---|---|---|
| Successful 275 applicants | 194 | 81 | 29.5% |
| Top 275 refereed applications | 211 | 64 | 23.3% |
| Top 275 publishing applicants | 218 | 57 | 20.7% |
| Top 275 cited applicants | 211 | 64 | 23.3% |

## 10. Conclusions and discussion

Research councils use different varieties of peer review, past performance measurements of applicants, and panel review methods when deciding which are the applicants and applications should be funded. The aim of this study was to compare in a specific case the peer-review based assessments of scientific quality with scientometric performance indicators, and to investigate the extent to which funding decisions are guided by the outcomes of the peer review process or past performance indicators. We have shown that this Council's procedures operate well in identifying and discarding the tail of the distribution. However, within the top half of the distribution, neither the review outcomes nor past performance measures correlate positively with the decisions of the Council.[16]

Obviously, the Council has a large autonomy in prioritizing the applications. In this process both external reviews and performance indicators play only an auxiliary role. As was shown, the referee reports — which are organized by the Council itself — are better predictors for obtaining grants than the indicators of past performance. However, the Council takes its own responsibility in the case of conflicting referee comments, or when its members' assessments differ from the opinions of the referees or the performance indicators. The same holds true in the case of gender bias.

The specific social mechanisms that dominate the decision-making process have to remain a subject for future research. Our findings suggest several directions to explore:

- The positive effects of the quality of co-applicants on the probability of getting funded suggest a network analysis of applicants, reviewers, committee members, and Council board members. Is funding correlated to the visibility of the applicants within these networks? This impression is reinforced by the priority of citations when compared with publications in the prediction using discriminant analysis.

- The distribution of funds over disciplines also seems to point to contextual factors influencing the grant allocation, especially in case of the large share of psychology;

- The positive gender correction corresponds to explicit policies in the case we studied.

---

16 At the discipline level the indicators work slightly better and the correlations are higher. This was expected. Peer review based indicators are more adequate on a low level of aggregation. However going from discipline to sub-discipline did not further improve the results.

Further research may also address the limitations of this study. First, we studied only the open competition and career programs of a single research council. Extending the study to other funding schemes (such as mission-oriented and thematic programs) and to other disciplines would be useful. Secondly, we did not study *post performance*, as this was not possible with the available data. An important question is the extent to which researchers funded by the Research Council perform better *ex post.* Thirdly, the definition of past performance takes into account only a part of the research output. For example, books and book chapters were not included in our analysis, although in some of the disciplines they are considered to be more important than articles. Thus, we emphasize that the conclusions of this study are valid only for those disciplines in which international journals are the dominant form of output (such as economics and psychology). Finally, the outcomes of the study may be country specific, and therefore comparison with cases in other countries would be useful.

In our opinion, other studies suggest that our results may be valid in more general terms. A study of some of the same funding instruments in other research fields in the Netherlands has shown the same differences that we found between the top and the tail of the distribution, without however studying differences within the top half of the distribution (Van Leeuwen 2007). Our hypothesis is that doing so would confirm our results also for fields such as the earth sciences and chemistry. Additionally, a detailed evaluation of the *Boeringer Ingelheim Fonds* by Bornmann & Daniel (2008) shows that even though the selected applicants are better on average than the non-selected applicants (as in our study), a large percentage of Type I errors (funded applicants who do not perform highly afterwards) and Type II errors (non-funded applicants who later become top researchers) were still detected. This means that here again, discrimination is modest at best. Our hypothesis would be that by using our approach of first correcting for the tails these authors might obtain results similar to ours.

A recent study by Böhmer *et al.* (2008, at pp. 115 ff.) applied precisely our method to applicants to the prestigious Emmy-Noether programs for post-doc funding in the German Federal Republic. The results varied among disciplines: in chemistry and biology the funded applicants were more prolific than the non-funded, yet this relation was the opposite for medicine. In physics, there was no difference between those who were funded and those who were not. The researchers note that their results are indeed counter-intuitive. We therefore hypothesize that the relationships found in this paper for the social sciences may hold true for other fields as well.

## 11. Policy implications

Our study suggests that picking the best at the individual level is hardly possible. In the social process of granting proposals many processes play a role, apart from scholarly quality: bias, old-boys' networks and other types of social networks, bureaucratic competencies, dominant paradigms, etc., all play an important role in selection processes. The policy implications therefore should go in the direction of improving the reflexivity of the procedures. The initiative taken by the Council to collaborate in this study may itself be an indication of this Council's growing awareness of the many issues at stake.

Given financial constraints, serious selection procedures are organized in order to choose the best proposals. However, criteria and indicators are never unambiguous, and this holds true not only for bibliometric indicators and peer review, but also for the (implicit) criteria used by decision makers. As a consequence, the quality of the grant allocation mechanism can be considered a *systems level issue* (Osterloh & Frey 2008). Instead of (fruitlessly) trying to improve procedures (and statistical indicators) for selecting individual projects, the main issue is to ensure that the system works properly despite uncertainties. The following relevant questions/ issues have emerged from our study:

- Quality requires both variation and selection. Does the funding system support the required variation (through a variety of funding institutions)? Is the selection process adequate (e.g., a variety of criteria; openness to innovation)? Are roles assigned adequately (e.g., neutrality of referees)?

- The discarding of the tail of the distribution can be done on the basis of more technocratic assessments (using bibliometrics) given the current criteria. This may be different if one wants to stimulate innovative research, because innovation is not necessarily correlated with these performance indicators; the latter are by their nature conservative. In other words, even if the procedure supports good (mainstream) research, it does not necessarily support innovation.

- Within the core (top) group, one should be more aware of the nature of the review process as a form of external organization and the potential bias that is thus implied. From the perspective of organization theory, the reviewers can be considered as an (external) instrument of the Council to reach a decision, and not as an independent assessment of and decision about the quality of the

> applications. This suggests the need for regularly evaluating a potential bias
> that creeps into the procedures, and regularly changing committees, panels,
> juries, and reviewers in order to keep the system open. The highly skewed
> distribution of grants over disciplines suggests that this may require periodic
> attention. To force the system to behave in this way, it may be useful to
> introduce competition between funding agencies.

In sum, we have found a negative answer to the Council's question of whether it selects the "best" researchers and proposals in terms of past performance and anonymous reviews.  At the same time, however, this counter-intuitive finding does not need to be considered a problem. From the perspective of science policy, the role of a Research Council is not only to select the best researchers, but to improve scientific research more generally, and to maximize the probability of scientific breakthroughs. Allocating funds to the researchers with *currently* the highest status can be considered only a specific (and to be evaluated) means to this end. The procedures for allocating funds can be discussed in terms of the effectiveness and efficiency of fulfilling this function at a systems level.[17] In addition to being very time-consuming, the current selection procedures are perhaps increasingly ineffective beyond removing the tails of the distribution. The system generates necessarily Type I and Type II errors, and these may harm the research system (e.g. in terms of decisions for tenure) more than can be made visible from evaluating the allocation mechanism itself.

**Bibliography**

Adler, Robert, John Ewing, & Peter Taylor, *Citation Statistics*. International Mathematical Union 2008

Aksnes, Dag W. & Randi Elisabeth Taxt, Peer review and bibliometric indicators: a comparative study at a Norwegian university. *Research Evaluation* **13** (2004) 33-41

Andersen, H., Editorial - journals under threat: a joint response from HSTM editors. *Metascience* **18** (2009) 1–4

Barker, Katherine, The UK Research Assessment Excercise: the evolution of a national research evaluation system. *Research Evaluation* **16** (2006) 3-12

---

[17] For example, one could argue that top researchers already have enough resources to perform outstandingly. From an economic point of view, it might then be advisable to stimulate the larger group of above-average researchers as a way of maximizing the marginal effects of the investments.  Doing so might help these researchers to improve their performance more easily than the already top-performing researchers. By using grants to support as many *good* researchers as possible, one might help to move the scientific landscape upward as a whole.

Böhmer, Susan, Stefan Hornbostel, & Michael Meuser, Postdocs in Deutschland: Evaluation des Emmy Noether-Programms. *iFQ-Working Paper* No. 3, May 2008; available at http://www.forschungsinfo.de/Publikationen/Download/working_paper_3_2008.pdf (last visited Sep. 16, 2008).

Bornman, Lutz & Hans-Dieter Daniel, Potential sources of bias in research fellowship assessments: effect of university prestige and field of study. *Research Evaluation* **15** (2006a) 209-219

Bornman, Lutz & Hans-Dieter Daniel, Selecting scientific excellence through committee peer review. *Scientometrics* **68** (2006b) 427-440

Bornman, Lutz & Hans-Dieter Daniel, Reliability, fairness, and predictive validity of the peer review process for the selection of research fellowship recipients of the Boehringer Ingelheim Fonds.In: Barbara M. Kehm (Ed.), *Hochschule im Wandel. Die Universität als Forschungsgegenstand. Festschrift für Ulrich Teichler* Frankfurt am Main: Campus, 2008a, 365-376.

Bornman, Lutz, Rudiger Mutz & Hans-Dieter Daniel, Gender differences in grant peer review, a meta-analysis. *Journal of Informatrics* 2008b

Butler, Linda, Assessing University research, a plea for a balanced approach. *Science & Public Policy* **34** (2007) 565-574.

Butler, Linda, Using a balanced approach to bibliometrics: quantitative performance measures in the Australian Research Quality Framework. *Ethics in Science and Environmental Politics* **8** (2008)

Charlton, Bruce & Peter Andras, Evaluating universities using simple scientometreic research-output metrics. *Science & Public Policy* **34** (2007) 555-563

Cicchetti, D. V. (1991). The reliability of peer review for manuscript and grant submissions: a cross-disciplinary investigation. *Behavioral and Brain Sciences,* 14(1), 119-135.

De Haan, Jos, *Research Groups in Dutch Sociology*. Amsterdam: Thesis Publishers, 1994.

Donovan, Clair, Introduction: future pathways for science policy and research assessment: metric vs peer review, quality vs impact. *Science & Public Policy* **34** (2007) 538-542

ESF, *Peer review: its present and future state*. Eurohorcs, 2006.

Frohlich, Gerhard, Anonyme Kritik: Peer review auf dem Prüfstand der Wissenschaftsforschung. *Medizin-Bibliothek-Information* **3** (2003) 33-39)

Hirschauer, Stefan, Peer Review Verfahren auf dem Prüfstand – zum Soziologiedefizit der Wissenschaftsevaluation. *Zeitschrift für Soziologie* **33** (2004) 62-83.

Langfeldt, Liv, Decision making constraints and processes of grant peer review, and their effect on review outcome. *Social Studies of Science* **31** (2001) 820-841

Langfeldt, Liv. Expert panels evaluating research: decision-making and sources of bias. *Research Evaluation* **13** (2004) 51-62.

Laudel, Grit, & Gloria Origgi, Special issue on the assessment of interdisciplinary research. *Research Evaluation* **15** (2006) 2-80

Lawrence, Peter A., The mismeasurement of science. *Current Biology* **17** (2007). Issue 15, 7 August, R583-R585

Leydesdorff, Loet, *Caveats* for the Use of Citation Indicators in Research and Journal Evaluations, *Journal of the American Society for Information Science and Technology* **59**(2) (2008) 278-287

Martin and Irvine (1984). *Foresight in Science: Picking the Winners.* London: Pinter.

Melin, Göran, and Rickard Danell, The top eight percent: development of approved and rejected applicants for a prestigious grant in Sweden. *Science & Public Policy* **33** (2006) 702-712

Moed, Henk F., The future of research evaluation rests with an intelligent combination of advanced metrics and transparent peer review. *Science & Public Policy* **34** (2007) 575-583

Mulkay, Michael J., The mediating role of the scientific elite. *Social Studies of Science* **6** (1976) 445-470.

Nederhof A.J. Bibliometric monitoring of research performance in the social sciences and the humanities: A review. *Scientometrics* **66** (2006) 81-100

Nightgale, Paul & Alister Scott, Peer review and the relevance gap. Ten suggestions for policy makers. *Science & Public Policy* **34** (2007) 543-553

Oppenheim, Charles, Out with the old and in with the new: the RAE, bibliometrics and the new REF. *Journal of Librarianship and Information Science* **40** (2008) 147-9

Osterloh, Margrit & Bruno S Frey, *Creativity and conformity: Alternatives to the present peer review system.* Paper *Peer Review Reviewed* Workshop. Berlin: WSB, 24-25 April 2008.

Peters, D.P. & Ceci, S.J. Peer-review practices of psychological journals: The fate of published articles, submitted again. *Behavioral and Brain Sciences* **5** (1982) 5(2): 187-195.

Rothwell, Peter M., & Christopher N. Martyn, Reproducibility of peer review in clinical neuroscience. *Brain* **123** (2000) 1964-1969.

Sandstrom, Ulf & Martin Hällsten, *Springboard or stumbling block — can research Councils promote scientific excellence without gender bias?* Swedish research Council 2005.

Sandstrom, Ulf & Martin Hällsten, Persistent nepotism in peer review. *Scientometrics* **74** 2008, 175-189

SWR-RGW, *Judging research on its merits.* Amsterdam: KNAW 2005

The British Academy, *Peer review: the challenges for the humanities and the social sciences.* London 2007

Van den Besselaar, Peter, *A map of media and communication research.* Den Haag, Rathenau Instituut, 2007. (SciSA rapport 0705 – in Dutch)

Van den Besselaar, Peter & Loet Leydesdorff, *Past performance as predictor of successful grant applications, a case study.* Den Haag: Rathenau Instituut, 2007. (SciSA rapport 0706)

Van Leeuwen, Thed N., *To which extent does NWO VI funding support excellent research?* Centre for Science and Technology Studies (CWTS), Leiden University, The Netherlands (2007)

Versleijen Anouschka, Barend van der Meulen, Jan van Steen, Robert Braam, Penny Boneschanker-Kloprogge, Ruth Mampuys, & Peter Van den Besselaar, *Thirty* *years public research funding in the Nederlands (1975-2005): trends and policy debates.* Den Haag, Rathenau Instituut: 2007. (SciSA rapport 0703 – in Dutch)

Wenneras, Christine & Agnes Wold, Nepotism and Sexism in Science. *Nature* **387** (1987) p341-343

Wessely, Simin, Peer review of grant applications: what do we know? *The Lancet* **352** (1998, July 25) 301-305